\documentclass[aps,showpacs,floats,superscriptaddress]{revtex4-2}

\usepackage{hyperref}
\usepackage{graphicx}
\usepackage{epstopdf}
\usepackage{amssymb}
\usepackage{amsmath,bm}
\usepackage{psfrag}
\usepackage{epsfig}
\usepackage{float}
\usepackage{bm}

\begin{document}

\title{Moir\'e-regulated composition evolution kinetics of bicomponent nanoclusters}

\author{Mikhail Khenner\footnote{Corresponding
author. E-mail: mikhail.khenner@wku.edu.}}
\affiliation{Department of Mathematics, Western Kentucky University, Bowling Green, KY 42101, USA}
\affiliation{Applied Physics Institute, Western Kentucky University, Bowling Green, KY 42101, USA}

\begin{abstract}
\noindent

A simple model and computation of Moir\'e-regulated composition evolution kinetics of bicomponent nanoclusters is presented.
Assuming continuous adsorbate coverage on top of 2D bilayer and Moir\'e potential-driven nanocluster 
formation at fcc sites of Moir\'e landscape, 
these sites experience the influx of one component of a bicomponent adsorbate and the outflux of another component.
Kinetics of this process is characterized for several combinations of adsorption potentials and their relative strengths.

\end{abstract}

\date{\today}
\maketitle


\section{Introduction}
\label{Intro}

Bilayers of 2D materials, such as twisted bilayer graphene, or a single-layer 2D material (e.g., graphene) deposited onto a high-symmetry 
surface of a noble metal, feature a biperiodic structure of hills and valleys known as Moir\'e superlattice. 
This structure gives rise to spatially selective adsorption of atoms or molecules and it also presents a complex landscape for surface diffusion. 
Therefore, Moir\'e-regulated self-assembly has been widely used to create ordered 2D arrays of a single-metallic nanoclusters and 
nanoparticles \cite{DBFM}-\cite{PLRMBK}. Less known is that bimetallic nanoclusters and nanoparticles also can be fabricated using this 
technique \cite{MWSBH}-\cite{DBSBSSP}. These authors prepared the substrate size (up to 1 cm$^2$) regular arrays of PdAu 
alloy nanoclusters on 
nanostructured ultrathin alumina films \cite{MWSBH}, PtRu alloy nanoclusters on graphene/Ru(0001) \cite{EBLB}, and PdPt alloy nanoclusters 
on graphene/Rh(111) \cite{DBSBSSP}. They demonstrated an independent control of the nanocluster size and its chemical composition by controlling 
the total amounts and the ratio of deposited metals. Also they demonstrated  
the narrow size distribution of the nanoclusters. The latter has been shown due to the regular distribution of the nucleation centers, 
which results in the uniform cluster growth rate. Notably, the metals may be deposited sequentially \cite{MWSBH,EBLB} or 
simultaneously \cite{DBSBSSP}. These studies pave the way to the studies of the reactivity of bimetallic clusters as a 
function of the composition at a given size, which is important for catalysis. Regular bimetallic nanocluster arrays serve as high-quality 
model nanocatalysts, allowing to study electrocatalytic oxidation reactions, sintering, poisoning, or to stabilize species like aldehydes. 
Notably, two fundamental electrocatalytic reactions in polymer
electrolyte membrane fuel cells are the oxidation of CO \cite{DBSBSSP} or
methanol. 

In this short communication we build upon our model of nanocluster self-assembly on graphene Moir\'e \cite{MyJAP_Moire} and our modeling of 
composition patterning in bimetallic surface alloys \cite{MySurfSci}-\cite{MySurfSci1}, to compute Moir\'e-regulated composition evolution kinetics  
of bicomponent nanoclusters. We deliberately keep the modeling generic, i.e. we do not assume a concrete bilayer substrate 
and a pair of metals that form a bimetallic nanocluster array. For proof of concept, we construct a primitive Moir\'e potential via the rotation (twist) of a 
top unstrained honeycomb lattice with respect to the identical unstrained bottom lattice. 
Next, we formulate evolution equation for the space and time-dependent composition of a bicomponent adsorbate that is driven by Moir\'e potential 
and, starting from a spatially uniform composition, quantify evolution kinetics of nanocluster composition.

\section{The Model}
\label{Model}

\subsection{Moir\'e potential}
\label{MoireModel}

\begin{figure}[H]
\vspace{-0.2cm}
\centering
\includegraphics[width=5in]{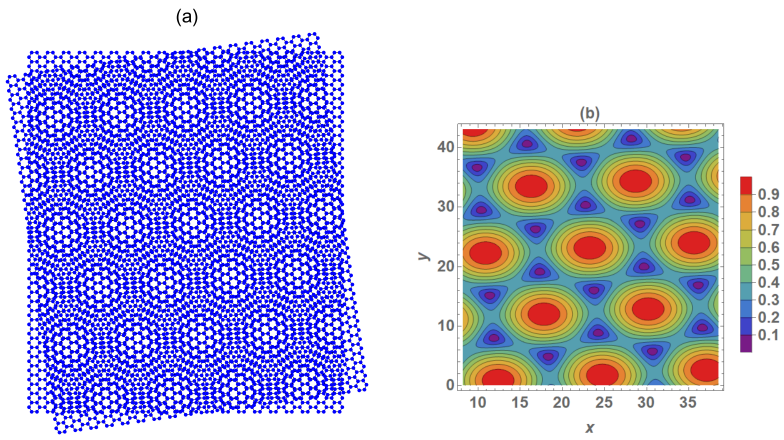}
\vspace{-0.15cm}
\caption{(a) Moir\'e formed by two identical hexagonal lattices rotated by $8^\circ$. (b) Corresponding Moir\'e potential $\epsilon_m(x,y)$. Red regions are top sites, blue regions are fcc sites.
}
\label{Fig1}
\end{figure}

Dimensionless Moir\'e potential is given by \cite{CWCF,WMCF,MyJAP_Moire}:
\begin{eqnarray}
\epsilon_m(x,y)&=&\frac{1}{3}\sqrt{3+2\left[\cos{\bm{x}\cdot \bm{c}_1}+\cos{\bm{x}\cdot \bm{c}_2}+\cos{\bm{x}\cdot \left(\bm{c}_1-\bm{c}_2\right)}\right]}, \label{Mpotent1}\\
\bm{c}_1&=&M^{-1}\left(1/2,\sqrt{3}/2\right)^T,\quad \bm{c}_2=M^{-1}\left(-1/2,\sqrt{3}/2\right)^T,\quad \bm{x}=(-\frac{4\pi}{3}x,\frac{4\pi}{3}y). 
\label{Mpotent2}
\end{eqnarray}
Here the (antisymmetric) scale matrix $M$ for unstrained 2D bilayer is:
\begin{equation}
M= 
\begin{pmatrix}
0 & \frac{1}{\theta}\\
-\frac{1}{\theta} & 0
\end{pmatrix},
\label{Am}
\end{equation}
where $\theta$ is the twist angle of a (honeycomb) top atomic lattice with respect to a (honeycomb) bottom atomic lattice in a bilayer. 
The lattices are assumed identical, i.e. they represent the same 2D material. The effect of the rotation is
the magnification of the bottom lattice and the emergence of a large-scale Moir\'e structure with a potential (\ref{Mpotent1}).  
Fig. \ref{Fig1}(b) shows Moir\'e potential that results from $8^\circ$ counter-clockwise rotation. 
The plotting window corresponds to the actual biperiodic domain used to compute Moir\'e-regulated composition evolution kinetics  
of a bicomponent adsorbate. Note that the computational domain contains five complete Moir\'e cells.
The unit of length in Fig. \ref{Fig1} is equal to graphene lattice spacing $a_g$.
Also we chose to have the minima of Moir\'e potential at fcc sites of Moir\'e lattice, and correspondingly, the maxima are 
at top sites, as this is the most common situation across bilayer substrates in experiment \cite{DBFM}-\cite{PLRMBK}. This choice implies 
nanocluster formation at fcc sites; the modeling aims to compute evolution kinetics of nanocluster composition at these sites.

\begin{figure}[H]
\vspace{-0.2cm}
\centering
\includegraphics[width=6.5in]{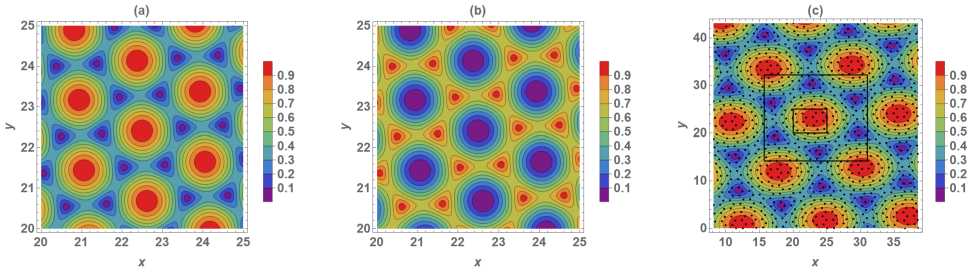}
\vspace{-0.15cm}
\caption{Adsorption potentials (a) $\epsilon_a^{(I)}(x,y)$ and (b) $\epsilon_a^{(II)}(x,y)$. (c) shows the 0.1 level set plot of
 $\epsilon_a^{(I)}(x,y)$ (the tiny black ovals resembling points) overlayed onto Moir\'e potential $\epsilon_m(x,y)$. 
 Smaller frame 
 corresponds to the plotting window of the panels (a) and (b).  Larger frame marks a single Moir\'e cell.
}
\label{Fig2}
\end{figure}

Adsorption potentials $\epsilon_a^{(I)}(x,y)$ and $\epsilon_a^{(II)}(x,y)$ of two atomic species at the top layer of a bilayer substrate are 
given by the expressions similar to Eq. (\ref{Mpotent2}):
\begin{eqnarray}
\epsilon_a^{(I)}(x,y)&=&\frac{1}{3}\sqrt{3+2\left[\cos{\bm{x}\cdot \bm{c}_1}+\cos{\bm{x}\cdot \bm{c}_2}+\cos{\bm{x}\cdot \left(\bm{c}_1-\bm{c}_2\right)}\right]}, \label{Mpotent1a}\\
\bm{c}_1&=&\left(1/2,\sqrt{3}/2\right)^T,\quad \bm{c}_2=\left(-1/2,\sqrt{3}/2\right)^T,\quad \bm{x}=(-\frac{4\pi}{3}x,\frac{4\pi}{3}y), 
\label{ads_pot1}
\end{eqnarray}
\begin{equation}
\label{ads_pot2}
\epsilon_a^{(II)}(x,y)=1-\epsilon_a^{(I)}(x,y).
\end{equation}
Notice that the scale matrix $M$ is absent from the vectors $\bm{c}_1$ and $\bm{c}_2$, since adsorption takes place on 
the top 2D layer of a bilayer substrate, whereby small atomic cells determine the adsorption sites. (Equivalently, $M=I$, where $I$ is 2$\times$2 identity matrix.) 
Figures \ref{Fig2}(a,b) show these adsorption potentials in a correspondingly smaller biperiodic plotting window. 
The first atomic species (either A or B) preferentially adsorbs at fcc sites on the top 2D layer of a bilayer substrate, whereas the second 
atomic species preferentially adsorbs at top sites.
For comparison of vastly 
different scales of Moir\'e and the adsorption potentials, Fig. \ref{Fig2}(c) shows the 0.1 level set plot of
 $\epsilon_a^{(I)}(x,y)$ overlayed onto Moir\'e potential $\epsilon_m(x,y)$ in the computational window. The 0.1 level set plot contains  
the minima of $\epsilon_a^{(I)}(x,y)$, i.e. it shows the locations of fcc adsorption sites on the top 2D layer of a bilayer substrate. 

\subsection{Evolution equation for Moir\'e-regulated composition of bicomponent nanoclusters}
\label{DiffEq}

In this section we formulate equation that models composition evolution kinetics of a bicomponent adsorbate 
(a binary mixture) on a Moir\'e formed by two mutually rotated 2D honeycomb atomic lattices. 
A Moir\'e of unstrained twisted bilayer graphene is used as example.

Let $\nu_A$ and $\nu_B$ be the surface densities, or the surface concentrations (compositions) of A and B adsorbate components
(shortened to A and B adsorbates in the rest of the paper)
immediately after deposition on the top of 2D bilayer. 
Let also 
$X_A(x,y,t)$ and $X_B(x,y,t)$ be the local time-dependent surface concentrations in the 
course of nanocluster directed self-assembly. Define the dimensionless local time-dependent concentrations as $C_A(x,y,t)=X_A/\nu_A$ and 
$C_B(x,y,t)=X_B/\nu_B$. Mass conservation requires $C_A(x,y,t)+C_B(x,y,t)=1$, and without significant loss of generality we assume $\nu_A=\nu_B=\nu$ 
to reduce the number of dimensionless parameters. 
We choose $a_g$ and $a_g^2 k T \nu/D_B\gamma_B$ for the length and time scales, respectively. Here $a_g$ is graphene lattice 
spacing, $k T$ is Boltzmann's factor, and $D_B$, $\gamma_B$ are the surface diffusivity and the surface 
energy of B adsorbate. 

According to classical irreversible thermodynamics the dimensionless evolution equation for the concentration of B adsorbate reads: 
\begin{equation}
\frac{\partial C_B}{\partial t} =   \bm{\nabla} \cdot \left(L_{AB} \bm{\nabla} \mu_A + 
L_{BB} \bm{\nabla} \mu_B\right),   \label{B-eq}
\end{equation}
where 
$L_{AB}$, $L_{BB}$ are the kinetic transport coefficients and $\mu_A$ and $\mu_B$ the chemical potentials.
The latter are given by \cite{RPA,ZVD}:
\begin{eqnarray}
\mu_A &=& -C_B\frac{\partial \gamma}{\partial C_B}+\xi_A \bm{\nabla}^2 C_B+\Omega \epsilon_m(x,y),
\label{chempot1}\\
\mu_B &=& \left(1-C_B\right)\frac{\partial \gamma}{\partial C_B}-\xi_B \bm{\nabla}^2 C_B+\Omega \epsilon_m(x,y), 
\label{chempot2}
\end{eqnarray}
where
\begin{equation}
\gamma = G \left(1-C_B\right) + C_B  + N\left[\left(1-C_B\right)\ln \left(1-C_B\right)+C_B\ln C_B+H C_B \left(1-C_B\right)\right]
\label{nondim_gamma}
\end{equation}
is the free energy \cite{RSN,SuoLu,LuKim,RPA}. In Eqs. (\ref{chempot1}), (\ref{chempot2}) $\Omega$ is the strength of Moir\'e potential. 
In Eq. (\ref{nondim_gamma}) the first two terms constitute the weighted surface energy of a bicomponent monolayer \cite{RSN}. Also, 
the first two terms in the bracket are the entropic contributions, and the last term in the bracket is the 
enthalpic contribution \cite{VB}.
Together, the three terms in the bracket are the regular solution model. All physical and dimensionless parameters are shown in Table \ref{T1}.

The kinetic transport coefficients are \cite{MyPRMat,RPA,MyJAP_Moire}:
\begin{equation}
L_{AB}=D \mbox{e}^{F_A\alpha(x,y)}C_B \left(1-C_B\right),\quad L_{BB}=\mbox{e}^{F_B\beta(x,y)}C_B^2, \label{LAB_LBB}
\end{equation}
where 
$\alpha(x,y)$, $\beta(x,y)$ are the placeholders for the adsorption potentials \cite{VK}. 
Thus $\alpha(x,y)$ or $\beta(x,y)$  can be assigned either $\epsilon_a^{(I)}(x,y)$ or $\epsilon_a^{(II)}(x,y)$. 
Notice that the potentials
$\epsilon_m(x,y)$, $\epsilon_a^{(I)}(x,y)$, and $\epsilon_a^{(II)}(x,y)$ have the same dimensionless depth one (see Fig. \ref{Fig2}); 
the strengths of the potentials are set by the parameters $\Omega$, $F_A$, and $F_B$. $F_A$ and $F_B$ are the inverses of the  
adsorbate-substrate binding energies. $D \mbox{e}^{F_A\alpha(x,y)}$ and $\mbox{e}^{F_B\beta(x,y)}$ are the diffusional mobilities 
of two atom species which comprise a bicomponent adsorbate. 
   
Eq. (\ref{B-eq}) is of an extended forced Cahn-Hilliard (CH) type. A typical value of the enthalpy $H$ (see Table \ref{T1}) precludes a thermodynamically driven phase 
separation (a spinodal instability) of a binary mixture. Thus changes in the composition of bicomponent nanoclusters
are expected to be caused by forced phase separation that is driven by Moir\'e potential.
Cahn-Hilliard gradient energy terms $\xi_A \bm{\nabla}^2 C_B$ and $-\xi_B \bm{\nabla}^2 C_B$ in the chemical potentials 
(\ref{chempot1}) and (\ref{chempot2}) account for adsorbate-adsorbate interactions \cite{MyJAP_Moire}.


%
\begin{table}[!ht]
\centering
{\scriptsize 
\begin{tabular}
{|c|c|}

\hline
				 
			\rule[-2mm]{0mm}{6mm} \textbf{Physical parameter and typical value}	 & \textbf{Dimensionless parameter and value} \\
			\hline
                        \hline
			\rule[-2mm]{0mm}{6mm} Graphene lattice spacing, $a_g=2.46\times 10^{-8}$ cm & \\
			\hline
			\rule[-2mm]{0mm}{6mm} Lattice spacing of A adsorbate, $a_A=3.92\times 10^{-8}$ cm  & \\
			\hline
			\rule[-2mm]{0mm}{6mm} Lattice spacing of B adsorbate, $a_B=2.7\times 10^{-8}$ cm  & \\
			\hline
			\rule[-2mm]{0mm}{6mm} Temperature, $T=20^\circ$ &  \\
			\hline
			    \rule[-2mm]{0mm}{6mm} Surface energy of A adsorbate, $\gamma_A=2.3\times 10^3$ erg$/$cm$^2$ \cite{VRSK}  &  \\ 
				\hline
				\rule[-2mm]{0mm}{6mm} Surface energy of B adsorbate, $\gamma_B=1.17\times 10^3$ erg$/$cm$^2$ \cite{VRSK}  &  Ratio of surface energies, $G=\gamma_A/\gamma_B=1.96$ \\
				\hline
				\rule[-2mm]{0mm}{6mm}  Surface density of either A or B adsorbate, $\nu=a_g^{-2}=1.65\times 10^{15}$ cm$^{-2}$  & Entropy, $N=k T \nu/\gamma_B=0.022$  \\
				\hline
				\rule[-2mm]{0mm}{6mm} Strength of Moir\'e potential, $\delta=1.2\times 10^{-11}$ erg  & Strength of Moir\'e potential, $\Omega= \delta \nu/\gamma_B=6.52$\\ 
			\hline
			\rule[-2mm]{0mm}{6mm} CH gradient energy coefficient, $\kappa=1.2\times 10^{-5}$ erg$/$cm  \cite{Hoyt} & CH gradient energy parameter, $\xi_A=\kappa a_A/a_g^2 \gamma_B=0.66$   \\
			\hline
			\rule[-2mm]{0mm}{6mm} & CH gradient energy parameter, $\xi_B=\kappa a_B/a_g^2 \gamma_B=0.46$ \\
			\hline
			\rule[-2mm]{0mm}{6mm}  Enthalpy of a binary mixture, $\eta=0.01 \gamma_B=11.7$ erg$/$cm$^2$ & Enthalpy, $H=\eta/k T \nu=0.45$ \\
			\hline
			\rule[-2mm]{0mm}{6mm} & Ratio of surface diffusivities, $D=D_A/D_B$ (varies)\\
			\hline

\end{tabular}}
\caption[\quad Parameters]{Physical and dimensionless parameters. $a_A$ and $a_B$ correspond to Pt and Ru, respectively. $D_A$, $D_B$, and the strengths of the adsorption potentials 
are not shown in the physical parameter column, since only the ratio $D=D_A/D_B$ and the 
dimensionless strengths $F_A$ and $F_B$ of the adsorption potentials are required. 
}
\label{T1}
\end{table}

\subsection{Results}
\label{Res}

Eq. (\ref{B-eq}) is computed using a Method of Lines framework. Spatial discretization on a fine 256x256 grid is via the fourth-order finite differences, 
the time integration is done by the implicit midpoint method. The logarithmic terms in the free energy are handled by the 
technique described in Ref. \cite{MySurfSci1}. The initial condition is $C_B(x,y)=0.1$ at $t=0$, i.e. a binary mixture has 10$\%$ of B atoms 
spread uniformly over Moir\'e landscape shown in Fig. \ref{Fig1}. 

We select six cases 1(a)-3(b) as follows:
\begin{enumerate}
\item $F_A=0$, $F_B=1.5$
\begin{enumerate}
\item $\beta(x,y)=\epsilon_a^{(I)}(x,y)$
\item $\beta(x,y)=\epsilon_a^{(II)}(x,y)$
\end{enumerate}
\item $F_A=1.5$, $F_B=0$
\begin{enumerate}
\item $\alpha(x,y)=\epsilon_a^{(I)}(x,y)$
\item $\alpha(x,y)=\epsilon_a^{(II)}(x,y)$
\end{enumerate}
\item $F_A=F_B=1.5$
\begin{enumerate}
\item $\alpha(x,y)=\epsilon_a^{(I)}(x,y)$, $\beta(x,y)=\epsilon_a^{(II)}(x,y)$
\item $\alpha(x,y)=\epsilon_a^{(II)}(x,y)$, $\beta(x,y)=\epsilon_a^{(I)}(x,y)$
\end{enumerate}
\end{enumerate} 
For each case we also vary the ratio of the surface diffusivities, $D$.
Note that the cases 1(a,b) correspond to A atom having equal probability to adsorb at fcc or top site on the top layer of a bilayer substrate, 
whereas the cases 2(a,b) correspond to B atom having equal probability to adsorb at fcc or top site. 
Value 1.5 for $F_A$ and $F_B$ is chosen because at larger $F_A$ or $F_B$ the excessive numerical stiffness makes the computation of 
Eq. (\ref{B-eq}) prohibitively slow.

\begin{figure}[H]
\vspace{-0.2cm}
\centering
\includegraphics[width=6.5in]{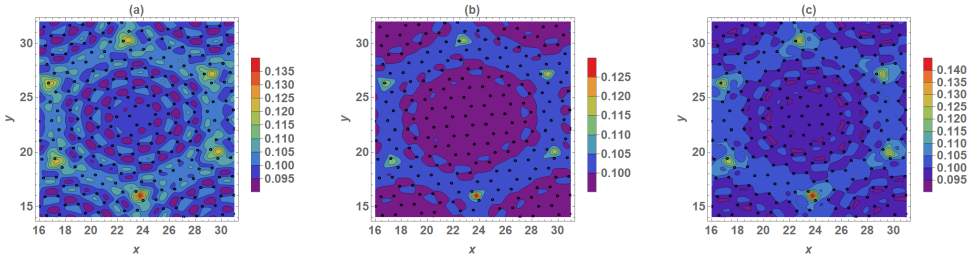}
\vspace{-0.15cm}
\caption{$C_B(x,y)$ at $t=0.02$ in the Moir\'e cell framed in Fig. \ref{Fig2}(c). $D=1$. (a) Case 1(a), (b) Case 2(a), (c) Case 3(a). The tiny black ovals show the 0.1 level set of 
$\epsilon_a^{(I)}(x,y)$.
}
\label{Fig3}
\end{figure}

As the example, for cases 1(a), 2(a), and 3(a) Fig. \ref{Fig3} shows the distribution of the concentration of B adsorbate in the central Moir\'e cell shortly after 
Moir\'e-regulated self-assembly 
has begun. Significant differences can be observed 
only due to 
differences in the mobilities of the adsorbate components that are caused by the choices of  
the adsorption potentials. Note that the maximum and minimum values of $C_B$ also differ.
Under the action of Moir\'e potential B adsorbate is driven out of the top site, i.e. out of the center of the Moir\'e cell 
and it starts to accumulate at fcc sites. 
With time, its concentration at fcc sites steadily increases, while the concentration at the top site steadily decreases. 
Correspondingly, the concentration of A adsorbate increases at the top site and decreases at fcc sites. Moir\'e-regulated self-assembly would 
terminate when  
B adsorbate depletes entirely at the top site ($C_B=0, C_A=1$) and correspondingly at fcc sites $C_B=1, C_A=0$. 

\begin{figure}[H]
\vspace{-0.2cm}
\centering
\includegraphics[width=6.5in]{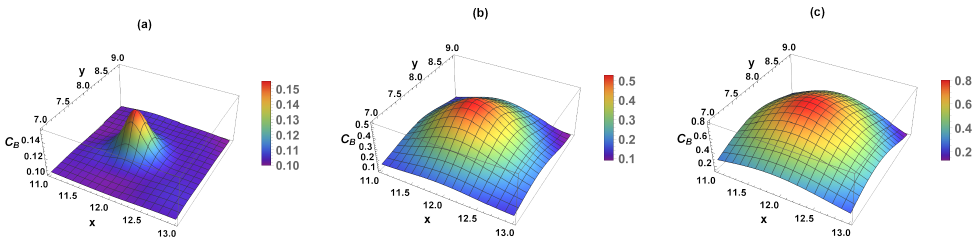}
\vspace{-0.15cm}
\caption{$C_B(x,y)$ across an fcc site for Case 2(a).  (a) $t=0.02$, (b) $t=0.5$, (c) $t=1$.
}
\label{Fig3a}
\end{figure}

$C_B$ increases at nearly the same rate at all fcc sites in the computational domain (Fig. \ref{Fig1}). In Fig. \ref{Fig3a} the typical spatial 
profile of $C_B$ over an fcc site is shown at three time instances. The profile broadens with time, whilst the maximum of $C_B$, attained at the site 
center, increases with time. 

\begin{figure}[H]
\vspace{-0.2cm}
\centering
\includegraphics[width=5in]{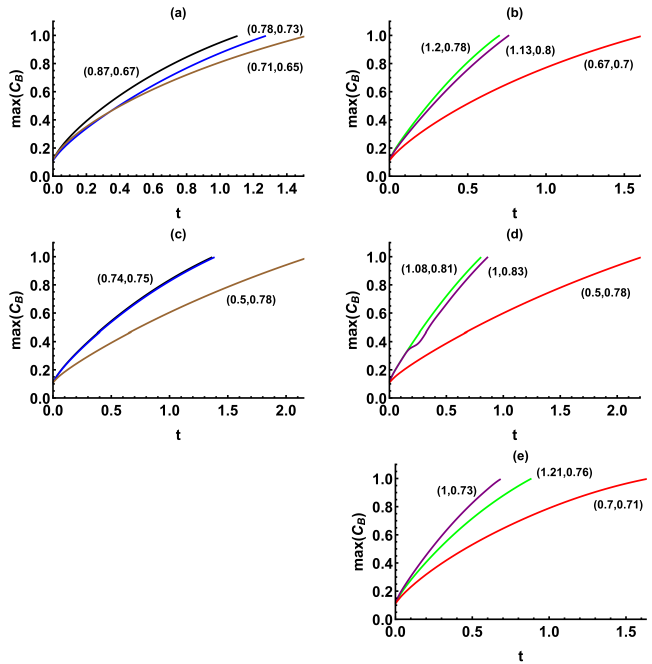}
\vspace{-0.15cm}
\caption{Maximum of $C_B(x,y)$ at an fcc site vs. time. (a) Case 3(a) (black curve), Case 1(a) (blue curve), Case 2(a) (brown curve) at $D=1$; (b) Case 3(b) (green curve), Case 1(b) (purple curve), 
Case 2(b) (red curve) at $D=1$; (c) Case 3(a) (black curve), Case 1(a) (blue curve), Case 2(a) (brown curve) at $D=0.1$; 
(d) Case 3(b) (green curve), Case 1(b) (purple curve), Case 2(b) (red curve) at $D=0.1$;
(e) Case 3(b) at $D=1.75$ (green curve), Case 1(b) at $D=2.5$ (purple curve), Case 2(b) at $D=1.1$ (red curve). All curves are the fits to data of the form
$\mbox{max}(C_B)=0.1+r t^z$; the pair $(r,z)$ is marked next to each curve.
}
\label{Fig4}
\end{figure}

To characterize the kinetics of nanocluster composition evolution, we choose one of fcc sites and in Fig. \ref{Fig4} plot $C_B$ value at the site center 
as function of time.
We observe that cases 3(b) and 1(b) provide the most efficient nanocluster enrichment by B component, with a weak dependence on $D$ 
(see the green and purple curves in Figures \ref{Fig4}(b), (d), (e)). 
For these cases $F_B=1.5$ and therefore the mobility of B adsorbate is large. 
Cases 2(a), 2(b) provide the least efficient nanocluster enrichment by B component,
with stronger dependence on $D$ (see Figures \ref{Fig4}(a)-(d)). 
For these cases $F_B=0$ and therefore transport of B adsorbate is slow. As $D$ decreases, also the mobility of A adsorbate decreases. 
This leads to weaker outflow of A component from nanoclusters, and therefore  
nanocluster enrichment by B component slows down (compare the brown curves in Figures \ref{Fig4}(a),(c) and the red curves in Figures \ref{Fig4}(b),(d)).

\section{Conclusions}

This paper presents a generic kinetic model for a continuous bicomponent adsorbate that is driven by Moir\'e potential. 
Moir\'e-regulated surface diffusion results in nanoclusters whose composition strongly depends on time and 
on the details and strength of the adsorption potentials of two atom species at the top layer of a 2D bilayer substrate. 
Typically, if the adsorbate-substrate 
binding energy of one adsorbate component is weak, then the corresponding diffusional mobility is strong, 
and a nanocluster 
becomes enriched by that component. Kinetics of such enrichment follows a power law in time, with the exponent in the 
range 0.7-0.83.   

\bigskip
\begin{center}
{\bf DECLARATIONS}
\end{center}

\medskip
\noindent
{\bf Ethical Approval}\\
\noindent
Not applicable.  \\

\medskip
\noindent
{\bf Funding}\\
\noindent
No funding was received for this research.


\end{document}